\documentclass[12pt,pra,aps,amssymb,amsfonts,amsmath,tightenlines]{revtex4}
\usepackage{dsfont}
\usepackage{graphicx}
\usepackage{amssymb,amsfonts,amsthm}
\usepackage{color}
\newcommand{\half}{\mbox{$\textstyle \frac{1}{2}$}}
\newcommand{\re}{\mbox{$\rm e$}}

\newcommand{\rd}{\mbox{$\rm d$}}
\begin{document}

\title{L\'evy Information and the Aggregation of Risk Aversion}

\author{Dorje C.~Brody$^*$, Lane~P.~Hughston$^\dagger$}

\affiliation{$^*$Mathematical Sciences, Brunel University, Uxbridge UB8 3PH, UK \\ 
$^\dagger$Department of Mathematics, University College London,
London WC1E 6BT, UK}

\date{\today}

\begin{abstract}
When investors have heterogeneous attitudes towards risk, it is reasonable to assume that each investor has a pricing kernel, and that these individual pricing kernels are aggregated to form a market pricing kernel. The various investors are then buyers or sellers depending on how their individual pricing kernels compare to that of the market. In Brownian-based models, we can represent such heterogeneous attitudes by letting the market price of risk be a random variable, the distribution of which corresponds to the variability of attitude across the market. If the flow of market information is determined by the movements of prices, then neither the Brownian driver nor the market price of risk are directly visible: the filtration is generated by an ``information process" given by a combination of the two. We show that the market pricing kernel is then given by the harmonic mean of the individual pricing kernels associated with the various market participants. Remarkably, with an appropriate definition of L\'evy information one draws the same conclusion in the case when asset prices can jump. As a consequence we are led to a rather general scheme for the management of investments in heterogeneous markets subject to jump risk. 

\begin{center}
{\scriptsize {\bf Keywords: Asset pricing, market information, heterogeneous markets, 
pricing kernel, \\  \vspace{-0.1cm}
risk aversion, risk premium, jump risk, signal processing, L\'evy information.
} }
\end{center}
%\subclass{MSC code\and JEL classification code}
\end{abstract}
\maketitle

%SECTION 1. Introduction
\section{Introduction}
\label{sec:I}

\noindent 
The importance of the flow of information in financial markets is clearly evident. As market 
participants we are all ``signal processors". It is logical therefore to base our financial 
models as much as reasonably possible on the information available to market participants, and 
to try to understand the way in which market signals are processed. One has to admit from 
the outset that markets are complicated, not only in terms of their structure, but 
also in terms of the investor psychology. Nevertheless, we shall show that there is scope for building relatively 
simple but intuitively natural mathematical models that capture the effects of information flows 
in heterogeneous markets, thus allowing one to address a variety of practical issues arising 
in the general area of investment management that might otherwise seem unapproachable.  

In a typical financial model it is usual to begin with a probability space $({\mathit\Omega}, 
{\mathcal F},{\mathbb P})$, together with a filtration $\{{\mathcal F}_t\}_{t \geq 0}$. 
The so-called ``physical'' probability measure $\mathbb P$ is meant to summarise the 
system of market probability assignments to various possible events, and the ``market 
filtration'' $\{{\mathcal F}_t\}$ is meant to summarise, for each time $t\geq 0$, the totality of 
information available to market participants up to time $t$. Clearly both of these ideas involve 
a good deal of idealisation, and it may be taking the notion of ``market efficiency'' too far to 
suppose that such a characterisation of the market is realistic. It makes better sense perhaps
to suggest that each market participant in some way implicitly builds their own version of the 
basic model, and then by some process all these different versions of the basic model are 
amalgamated to produce an overall effective model that represents the market. 
The individual investor then trades in a way that is consistent with the relation of their 
``private'' model to that of the market as a whole. 

Our purpose here is to examine a particular example of such a scheme, arising in connection 
with the apparent variability of opinion one observes concerning the expected rates of returns 
on financial assets. It is plainly obvious to the investment management community that 
well-informed, intelligent market participants will often have significantly differing opinions on 
rates of return---indeed, not only with respect to their views on the expected returns associated 
with individual assets, but also the expected returns associated with the market as a whole. 
Clearly some sort of hypothesis of ``natural variation'' is needed to model such a situation. 
The question thus arising is the following: how one can reconcile such a view with those well-established 
techniques of financial engineering that entail some sort of equilibrium, or, at least, absence of 
arbitrage, as part of the very basis of the modelling framework, usually coupled implicitly 
with the assumption of a high degree of homogeneity and uniformity across the market---some 
version of the \textit{law of one price}$\,$?  

Our approach will be to model the excess rate of return (above the interest rate) as a random 
variable, the interpretation of which reflects the spread of opinion in the market about the rate of return  
to be demanded in exchange for the assumption of a given level of risk. We shall assume that the investors 
can be modelled as having a degree of rationality, in the sense that they 
recognise that other investors have differing opinions, and that the collective effect of these 
differing opinions will have an effect on market movements. 

In the context of financial modelling, 
we take the accepted ``modern'' view that financial models are  by their nature 
ephemeral. By that, we mean that models are always used ``in the present'' for 
a specific purpose---pricing, hedging, asset allocation, decision making. Once the model 
has been used, then it is (so to speak) thrown away, and another model is 
constructed for the next task. The ``new'' model may  be identical in 
structure to its predecessor---perhaps differing only slightly in the assignment of some 
parameter values. Nevertheless, it is different: one starts each day (or minute, 
or microsecond) with a fresh model. 
In practice, the term ``new model'' is usually applied only if there is some significant structural difference involved---for example, in the sense that the Vasicek and Cox-Ingersoll-Ross interest 
rate models are structurally different. In casual discussion one would not normally say that two different versions of 
the Vasicek model with different values of the mean reversion parameter were distinct models. 
But it is useful to maintain the idea that one really is in fact working with different models---perhaps 
a better choice of words would be to say that one is working with a parametric family of models. 
Then the periodic adjustment of the parameters is the ``calibration'' of the model. Indeed, the pervasive need for a regular regimen of robust model recalibration is a dominant feature of much of modern banking.

How does one use a financial model? That depends on the particular type 
of problem one is trying to solve (pricing, hedging, asset allocation, and so on), but typically one uses 
the freshly calibrated model to generate (by simulation, or numerical integration, or exact solution) the 
trajectories of the asset prices under different outcomes of chance; and then certain functions 
of the trajectories are averaged, with appropriate weightings, to provide the figures needed 
for the particular application. 

The point is that when ``averaging'' is carried out in a financial model, one is typically
averaging over the outcome of chance in two different senses simultaneously---the first being the 
usual sense of the development of the trajectory as time goes by (for example, one averages 
over a multitude of distinct random walks); and the second being the sense that one averages 
over different views or characteristics of a multiplicity of market participants. Thus if one models 
the excess rate of return as a random variable,  the ``randomness'' of the excess rate of return is not necessarily to 
be interpreted---in the model---exclusively in the sense that one particular value turns out to be the ``correct'' one selected 
by chance (as if by coin flip), but rather in the sense that if one were to select an investor at 
random then one could say with what probability that investor will have a view or characteristic that lies in a 
certain range. Thus in the model it is the collective effect (via the weighted average) of variation in the future trajectory that determines the solution to a problem posed in the present.

In this paper we make use of pricing kernel methods, which turn out to be particularly 
useful, allowing one to distinguish between pricing issues 
and hedging issues. See Cochrane (2005) for an informal but comprehensive 
introduction to the application of pricing kernels in finance. The study of heterogeneous markets 
is still, one could probably say, in its infancy, and in a state of active development. Indeed, one can be overwhelmed by trying to contend with all the different types of heterogeneity that can arise in financial markets---heterogeneity in risk attitude, in impatience, in probability assignment,  in transmission of information, in network connectivity, in information processing speed, and so on. See Brown \& Rogers (2012), Duffie (2012), Ziegler (2003), and references cited therein, for overviews of some of the issues connected to heterogeneity in financial markets currently being pursued.  There is also a large literature devoted to portfolio management under partial information (see, e.g., Bj\"ork, Davis \& Land\'en 2010, and references cited therein).  
Our approach in what follows is 
novel inasmuch as it combines pricing-kernel methods with information-based pricing and 
elements of behavioural finance in an intuitive yet mathematically rigorous treatment aimed at 
problems of asset allocation and investment decision, with a view particularly on how to manage such problems
in the face of issues involving jumps in asset prices, in situations where the ``hedging paradigm'' for derivative pricing breaks down, and in the context of a post-crisis world-view where buy-side concerns are taken as seriously as sell-side concerns. 

The structure of the paper is as follows. In Section~\ref{sec:2} we consider the problem of a heterogeneous market in which investors have variegated attitudes toward risk. We specialise to the case in which asset prices are driven by Brownian motion, and we model the variation in attitude toward risk by taking the excess rate of return to be a random variable. We introduce the idea of an ``information process'' as the generator of the market filtration, and derive the conditional distribution of the market price of risk.  In Section~\ref{sec:3} we work out the form that the pricing kernel takes in such a model and derive the remarkable result that the market pricing kernel is given by the harmonic mean of the pricing kernels attributable to the various market participants based on their attitudes towards risk. In Section~\ref{sec:4} we work out the dynamics of a typical financial asset under the assumptions that we have made, and show how the dynamics can be represented in a way that is explicitly consistent with the absence of arbitrage.  In particular, we are able to show the existence of a Brownian motion adapted to the filtration generated by the information process and such that the dynamical equations of both the asset and the pricing kernel are both are driven by this ``market'' Brownian motion. Finally, in Sections~\ref{sec:5} and \ref{sec:6} we show how the general framework that we have considered in the case of Brownian motion based models can be extended very naturally to a wide family of models admitting jumps. 

%%%%%%%%%%%%%%%%%%%%%
%SECTION 2. Random Risk Aversion and Market Heterogeneity
%%%%%%%%%%%%%%%%%%%%%
\section{Random Risk Aversion and Market Heterogeneity}
\label{sec:2}

\noindent
It will be useful if one regards the market filtration $\{{\mathcal F}_t\}$ as being generated by a set of one or more ``information processes". By an information process we mean a process that carries noisy or imperfect information about some quantity that is of relevance to market 
participants. The notion is a quite general one, and a number of different situations arise in 
which one can model the flow of information relevant to the formation of prices. Examples 
include information flows concerning the market factors that determine dividends on stocks, defaults 
on bonds, or claims on insurance contracts. The approach that we are adopting is 
that of ``information-based asset pricing'', as represented in Brody, Hughston \& Macrina (2007, 
2008a,b, 2010), Brody, Davis, Friedman \& Hughston (2009), Brody \& Friedman (2009), 
Filipovic, Hughston \& Macrina (2011), Hoyle (2010), Hoyle, Hughston \& Macrina (2011), 
Hughston \& Macrina (2008, 2012), Macrina (2006), Macrina \& Parbhoo (2011), and 
Rutkowski \& Yu (2007). An important advantage of thinking of the filtration as being generated by information processes is that the treatment of informationally heterogeneous markets can then be pursued in a relatively straightforward way. We consider a set of information processes, some of which are accessible to investor A, some to investor B, some to investor C, and so on, generally with some overlap. If the overlap is substantial for a relatively large number of investors, then we can for some purposes call this the ``market filtration''. Generally speaking, the system of filtrations has a kind of hierarchical structure that takes the form of a lattice. In what follows, we shall work with a single market filtration, since our main concern is with heterogeneous attitudes towards risk rather than heterogeneous information flows, but the setup will be structured in such a way that the consideration of heterogeneous information flows is also feasible.

An interesting and important example of an information process arises rather naturally in the 
context of geometric Brownian motion (GBM) models when we try to generalise such models 
to situations where the rate of return on the stock is not known exactly. The GBM models are, 
needless to say, a little 
too simple as such to be taken seriously as real-world models for asset price dynamics. 
Nevertheless, they do capture rather succinctly certain key elements of the relation between risk 
and return, and it is in that context that we confine the discussion initially to the GBM 
class. The idea is that once one obtains some understanding of how to deal with random rates 
of return in the case of constant-parameter GBM models, then one might focus on how to 
generalise the modelling framework to incorporate more realistic features, such as stochastic volatility or the inclusion of jumps.  

We begin by consideration of the case of a single risky asset in the standard GBM family of models. 
The discussion that follows readily generalises to the situation where a number of risky assets are traded. For simplicity we present the case of a single such asset, and we assume that no dividends are paid over the time horizon considered. For the price we 
write
\begin{eqnarray}
S_t = S_0\, \re^{(r+\sigma\lambda)t} \re^{\sigma B_t - \frac{1}{2}\sigma^2t},
\label{price process}
\end{eqnarray}
where $S_0$ is the initial price, $\{B_t\}_{t\geq 0}$ is a standard Brownian motion, $r>0$ is the 
interest rate, $\sigma>0$ is the volatility, and $\lambda >0$ is the risk aversion factor. The 
term $\sigma\lambda$ is called the ``risk premium'' or ``excess rate of return''. For a 
fixed level of risk aversion, the risk premium increases if one increases the level of riskiness 
(as represented by the volatility), and for a fixed level of riskiness, the risk premium increases 
if one increases the level of risk aversion. Since $\lambda\sigma$ is linear in each factor, it 
follows that $\lambda$ has the interpretation of being the ``excess rate of return per unit of 
risk'', or ``market price of risk''  in the GBM model. It should be evident, on the other hand, that 
there is no \textit{a priori} reason why the excess rate of return should be bilinear. In fact, the 
case of a bilinear risk premium is quite special. For example, in a general L\'evy model the excess rate of 
return is a nonlinear function of the risk aversion and the volatility (Brody, Hughston \& Mackie 
2012, Mackie 2012), and the notion of ``market price of risk'' is inappropriate. 

To complete the specification of the model we need a pricing kernel $\{\pi_t\}_{t>0}$, which in 
the standard GBM model takes the form
\begin{eqnarray}
\pi_t = \re^{-rt} \re^{-\lambda B_t - \frac{1}{2}\lambda^2t}. 
\label{pricing kernel}
\end{eqnarray}
The pricing kernel in an arbitrage-free model has the property that its product with the 
price of any non-dividend-paying asset gives a martingale under the physical measure 
${\mathbb P}$.  In the present situation we have
\begin{eqnarray}
\pi_tS_t = S_0 \,\re^{(\sigma-\lambda) B_t - \frac{1}{2}(\sigma-\lambda)^2t}, 
\end{eqnarray}
and one sees that the martingale condition is indeed satisfied. It is important to observe that for the expression of the principle of no arbitrage one needs to specify both the asset price and the pricing kernel. 

Thus, in summary, in the case of a single risky asset (and under the assumption that no 
dividends are paid over the time horizon considered), the model is given by the price process 
(\ref{price process}) and the pricing kernel (\ref{pricing kernel}). These processes are defined 
on a probability space $({\mathit\Omega},{\mathcal F},{\mathbb P})$ with respect to which 
$\{B_t\}_{t\geq 0}$ is a standard Brownian motion, and we can take the market filtration 
$\{{\mathcal F}_t\}$ as being the standard augmented filtration generated by $\{B_t\}$. The 
parameters of the model are $S_0$, $r$, $\lambda$, and $\sigma$. Once specified, the model 
can be used at time 0 to value and risk-manage certain other classes of asset. For example, 
if $H_T$ represents a random cash flow (perhaps the payoff of an investment strategy) at time 
$T$ determined by the trajectory of ${\{S_t\}}_{0 \leq t \leq T}$ over the time interval $[0,T]$,  
then in the model constructed at time $0$, the random value $H_t$, at any time $t \geq 0$, of the 
asset that delivers $H_T$ at time $T$ is represented by
\begin{eqnarray}
H_t =  \frac{1}{\pi_t}  {\mathds 1} ( t < T)\, \mathbb{E}_t [\pi_T H_T]\, ,
\end{eqnarray}
where ${\mathds 1} (\,\cdot \,)$ denotes the indicator function, and $\mathbb {E}_t[\,\cdot \,]$ 
denotes conditional expectation with respect to $\mathcal {F}_t$. For example, if 
$H_T = \max (S_t - K, \,0)$, then a calculation shows that $H_t$ is given by the familiar 
Black-Scholes formula for the value at time $t$ of a call option with strike $K$ and maturity 
$T$. Note that the pricing kernel methodology gives this result rather directly, 
without the involvement of hedging arguments, replication portfolios, market completeness, 
risk neutrality, change of measure, or the solution of partial differential equations---all of which, 
important and useful as they are in various specific contexts, are ultimately irrelevant to the 
determination of the price of an option once the pricing kernel has been specified. 

As another example, consider the optimal investment problem for an investor with utility 
$U(x)$ and initial endowment $H_0$, who wishes to invest in such a way as to maximise 
the expected utility of a contract that pays $H_T$ at time $T$. Assuming that $U(x)$, $x>0$, 
is a standard utility function satisfying $U'(x)>0$ and $U''(x)<0$, and writing $I(y)$, $y>0$, 
for the inverse marginal utility satisfying $I(U'(x)) = x$ for all $x>0$, then a variational 
argument shows that the optimal investment is in a contract that at time $T$ pays 
$H_T = I(\beta \pi_T)$, where the parameter $\beta$ is the (unique) solution to the budget 
constraint $H_0 = \mathbb{E} [\pi_TI(\beta \pi_T)]$. In the case of logarithmic utility 
$U(x) = \ln x$ , for instance, one finds that $H_T = H_0 / \pi_T$. Such results follow more 
or less directly from the pricing kernel methodology, without the need for consideration, for 
example, of the optimal portfolio strategy (if such exists) that will generate $H_T$. In 
principle, the investor simply pays $H_0$ and buys the contract that delivers $H_T$, and it is up to the seller 
whether they prefer (a) to accept the unhedged risk of delivering the contracted payment $H_T$ at $T$, or (b) to 
hedge the risk by using $H_0$ to construct a portfolio that is then managed in such a way 
as to produce the required $H_T$ at $T$.  

One sees that derivative contracts and investment management contracts are much 
the same thing from the point of view of the investor, at least if we add the further provision 
that the derivatives should have nonnegative payoffs. Perhaps the investment management 
paradigm has the moral advantage that at least in some sense the investor is clearly being 
sold a product that is optimal. Whether such well-defined optimisation criteria enter into the 
actual decision-making processes involved  in the marketing of investment opportunities and 
the targeting of clients is another matter---but clearly they should, to the extent that this is
practically possible, if we may speak normatively, and the same goes for the marketing of 
investment-grade derivatives. A key point is that the optimal investment plan typically involves 
characteristics of the investor (as modelled, for example, with the specification of a utility function),
together with the pricing kernel---but the microstructure of the market, as represented by the various
stocks that are traded, and so forth, does not come into play. 

Now suppose that the risk aversion factor (or excess rate of return per unit of risk) is not 
directly observable, and that there is uncertainty in the market as to its value. 
This state of affairs, as we have argued earlier,  is in many ways representative of reality,
and suggests a simple generalisation of the GBM model. Let us therefore write 
$X$ for the unknown value of the risk aversion factor, which we shall treat as a random variable. 
Then in our model for the typical asset price (assuming, for simplicity, that $X$ and $\{B_t\}$ are independent, that $X$ is positive, and that the other model parameters are constants) we have
\begin{eqnarray}
S_t = S_0\, \re^{(r+\sigma X)t} \re^{\sigma B_t - \frac{1}{2}\sigma^2t}.
\end{eqnarray}
Thus if we introduce a so-called  ``information process'' $\{\xi_t\}_{t \geq 0}$ defined by
\begin{eqnarray}
\xi_t = B_t  + Xt,
\label{information process}
\end{eqnarray}
we can write the price in the form
\begin{eqnarray}
S_t = S_0\, \re^{rt} \re^{\sigma \xi_t  - \frac{1}{2}\sigma^2t}.
\label{asset price xi}
\end{eqnarray}
Note that the asset price is monotonic in the information. It follows that the filtration 
generated by the asset price is the same as the filtration generated by the information 
process. Therefore, in our model it is rather natural to let this be the market filtration 
$\{{\mathcal F}_t\}$. Then the ``true'' value of the market risk aversion factor $X$ remains 
hidden, and at best can only be estimated by observations of the asset price (or, equivalently, 
the information). This is a rather satisfactory way of viewing the market, since it conforms to 
intuition, and allows for an embodiment of the idea that past performance is not 
necessarily a reliable guide to future performance. The information process has the property that
for large $t$ the value of $X$ is revealed. In particular, we have
\begin{eqnarray}
\lim_{t\to\infty} \frac {1}{t} \xi_t = X.
\end{eqnarray}
This relation follows from the fact that Brownian motion grows in magnitude like the square root of $t$. 
Thus investors do not know in advance the excess rate of return on an asset, but in the long run this is revealed. 
 
By use of the Bayes law, taking advantage of the fact that the random variable $\xi_t$ is conditionally Gaussian 
given $X$, one can work out the conditional distribution of the market factor $X$ given the 
relevant market information up to time $t$. The details of the calculation leading to this result are shown in Appendix~\ref{app:1}. One finds that
\begin{eqnarray}
p_t(\rd x)= \frac{\exp\left[
x\xi_t-\tfrac{1}{2} x^2
t\right]p(\rd x)}{\int^{\infty}_0 \exp\left[z\xi_t-\tfrac{1}{2}z^2 t\right]p(\rd z)},
\label{conditional distribution}
\end{eqnarray}
where the measure $p(\rd x)$ determines the unconditional  distribution of 
$X$.
% by the relation
%\begin{eqnarray}
%{\mathbb P} (X< x) = \int^{x}_0 p(\rd z).
%\end{eqnarray}
The conditional distribution of the risk aversion factor is then given by
\begin{eqnarray}
{\mathbb P} (X< x \, \left | \,{\mathcal F}_t \right) = \int^{x}_0 p_t(\rd z),
\end{eqnarray}
with $p_t(\rd x)$ as in (\ref{conditional distribution}), and it follows in particular that its conditional mean is 
\begin{eqnarray}
{\mathbb E}\, [X  \left | \,{\mathcal F}_t \right] = \int^{\infty}_0 x \, p_t(\rd x) \,.
\label{conditional expectation}
\end{eqnarray}

The statement ``$X$ is unknown'' can be interpreted in several ways. One is that there 
is a ``secret'' value of $X$ which none of the market participants know but the asset 
somehow  ``knows'', and that over time this secret value of $X$ works its way through the 
dynamics of the asset price to contribute to the eventual return displayed by the asset. Many 
people like to think in this way, even if they do not actually believe the asset ``knows'' 
anything. It is as though the market somehow  ``knows''. One sees this manner of thinking in the use of animistic language, in phrases like 
``the market is always right'', and also in the language of technical analysis. 
Another interpretation of ``X is unknown" is that there is variability of opinion in the market 
about the rate of return that ought to be expected for a given level of risk, and that the 
distribution of $X$ represents this spread of opinion.  Such variation might well 
be elemental, in the sense that each participant has their own private level of risk 
aversion, and that the distribution of $X$ reflects this. Indeed, it is a matter of human nature 
that equally intelligent and well-informed individuals can and will, by choice or disposition, 
exhibit markedly differing levels of risk tolerance and risk aversion. This is a practical fact of 
life that one encounters constantly in day-to-day interactions with other people (or for that matter animals). We know from experience that even 
a single individual can, depending on mood and circumstance, exhibit significantly variable 
attitudes towards risk. It seems therefore both necessary 
and reasonable to suppose that an equilibrium can be established in a market where investors 
have widely differing attitudes towards risk, and that market prices are obtained by averaging 
in some sense over all these different attitudes. 

Is it possible to reconcile the ``$X$ is secret'' point of view with the ``$X$ represents 
variation'' idea? From a modelling perspective, it would appear so. In particular, to 
calculate prices, one needs to form weighted averages over a large number of trajectories. 
One can imagine that each trajectory invoked in the averaging procedure involves some 
specific ``secret'' value of $X$; or alternatively, one can think of averaging over the whole 
market, taking into account all of the various risk preferences: the result is the same. 

%%%%%%%%%%%%%%%%%%%
%Section 3. Modeling the pricing kernel
%%%%%%%%%%%%%%%%%%%
\section{Modelling the pricing kernel}
\label{sec:3}

\noindent
With these thoughts in mind, we need to consider how to model the pricing kernel in a situation 
where heterogeneous attitudes towards risk prevail. One might be inclined simply to replace the 
parameter $\lambda$ in the GBW pricing kernel (\ref{pricing kernel}) with the random variable 
$X$ to give a tentative expression of the form 
\begin{eqnarray}
\pi_t  \stackrel{?}{=} \re^{-rt} \re^{-X B_t - \frac{1}{2}X^2t}
\end{eqnarray}
as a candidate for the pricing kernel. Unfortunately, this will not quite work, since once we 
introduce $\xi_t$ we obtain
\begin{eqnarray}
\pi_t  \stackrel{?}{=} \re^{-rt} \re^{-X \xi_t + \frac{1}{2}X^2t},  
\label{eq:14}
\end{eqnarray}
which is clearly not ${\mathcal F}_t$-measurable, on account of the explicit appearance of 
$X$, and as a consequence the associated process $\{\pi_t\}$ is not adapted to the filtration 
$\{{\mathcal F}_t\}$ generated by $\{\xi_t\}$. However, if we take the conditional expectation 
of the expression above with respect to ${\mathcal F}_t$, this gives a better candidate for the 
pricing kernel, namely: 
\begin{eqnarray}
\pi_t = \mathbb{E}_t \big[\exp\left( -rt  -X \xi_t + \half X^2t \right) \big] =  
\int^{\infty}_0  \exp\left( -rt  -x \xi_t + \half x^2t \right) \, p_t(\rd x),
\end{eqnarray}
which has the virtue of being ${\mathcal F}_t$-measurable. 
Then after insertion of expression (\ref{conditional distribution}) for $p_t(\rd x)$ we obtain
\begin{eqnarray}
\pi_t = \frac{1}{\int^{\infty}_0 \exp\left( rt + x\xi_t-\tfrac{1}{2}x^2 t\right)p(\rd x)}.
\label{pricing kernel xi}
\end{eqnarray}

This formula is perhaps most easily understood as follows. The process $\{n_t\}$ defined 
in terms of the pricing kernel by $n_t = 1/\pi_t$ is the so-called ``natural numeraire'' or 
``benchmark process'' (see, e.g., Long 1990, Flesaker \& Hughston 1998). The price 
process of any non-dividend paying asset, when expressed in units of the natural 
numeraire, is a martingale in the market filtration. Thus in the present context we have:
\begin{eqnarray}
n_t = \int^{\infty}_0 \exp\left(rt + x\xi_t-\tfrac{1}{2}x^2 t\right) p(\rd x).
\label{natural numeraire}
\end{eqnarray}

We note that for each value of $x$ the integrand corresponds to an asset with unit initial 
price and of the form (\ref{asset price xi}) with volatility $x$. Therefore, in the case of an 
unknown risk aversion factor we form a weighted portfolio of the numeraire assets 
obtained for various specific values of the risk aversion, and then invert this to obtain the 
pricing kernel $\pi_t = 1/n_t$. This leads us to the following important conclusion:~\textit{the 
pricing kernel associated with random risk aversion is given by the harmonic mean of 
the pricing kernels arising for various specific values of the risk aversion}. 

We have thus obtained a nice example of the use of information processes in shedding 
light on a problem of considerable interest in the investment community. Indeed, in the 
behavioural finance literature (see, for example, Shefrin 2008, 2009, and works cited 
therein), a good deal of evidence 
has been gathered to the effect that the correct way of amalgamating risk aversion in a 
heterogeneous market is not by simply averaging the risk aversion parameter over the market, 
but rather by taking a suitable average (typically a H\"older mean) of the associated stochastic 
discount factors. Thus if agent A says that the market price of risk should be $x$, and agent B 
says with equal conviction that the market price of risk should be $y$, then instead of naively 
averaging these numbers to obtain $\half (x+y)$ and inserting this figure into the stochastic 
discount factor to express the aggregate view, the behaviouralists propose first to work out the 
stochastic discount factors corresponding separately to the views of A and B, and then to take 
a suitable average. As we have seen above, our calculations support this general line of argument, 
and indeed we are able to go further by deducing from first principles a specific rule for the aggregation of risk 
aversion. 

%%%%%%%%%%%%%%%%%
%Section 4. information-based estimation of risk aversion
%%%%%%%%%%%%%%%%%
\section{information-based estimation of market risk aversion}
\label{sec:4}

\noindent
The following question can be posed. Is it possible to formulate the price dynamics of the 
risky asset in such a way that the resulting representations for $\{S_t\}$ and 
$\{\pi_t\}$  are expressible entirely in the language of stochastic differential equations, without 
explicit reference to the ``hidden'' risk aversion  variable $X\,$? By doing so, we would 
have a model formulated, so to speak, in the spirit of ``classical mathematical finance''. That 
is to say, the model would be proposed  in the form of a closed system of dynamical equations 
satisfied by the asset and the pricing kernel, with appropriate initial conditions, and some 
parametric freedom. Only later one would discover, as it were, that the solution to this 
system of equations implies and admits the existence of the hidden variables $X$ and $\{B_t\}$. 

It turns out that such a program is feasible, and indeed is rather enlightening, since it allows 
one to put forward a version of the theory described in the previous sections without the 
introduction of ``unobservable" elements, and yet with exactly the same practical conclusions. 
Furthermore, at the same time we are able to ``deduce'' the existence of a random variable 
$X$ having the characteristics already discussed, along with the associated ``true'' noise $\{B_t\}$, thus allowing the theory to admit the 
interpretation we have given it. 

We proceed as follows. As in the previous sections, we fix a probability space 
$({\mathit\Omega},{\mathcal F},{\mathbb P})$ and introduce a Brownian motion $\{B_t\}$ 
and an 
independent random variable $X$. We introduce the information process $\{\xi_t\}$ defined 
by (\ref{information process}), along with the filtration $\{{\mathcal F}_t\}$ that it generates, 
and we define the asset price by (\ref{asset price xi}) and the pricing kernel by
(\ref{pricing kernel xi}). By virtue of the relation $\rd \xi_t^{\,2} = \rd t$, the dynamical 
equation satisfied by the asset price takes the form
\begin{eqnarray}
\rd S_t = r S_t \,\rd t + \sigma S_t \, \rd \xi_t.
\end{eqnarray}
Our goal is to write the dynamics in a way that brings out more explicitly the fact that the 
price movements are being driven by Brownian motion. The only difficulty is that the Brownian 
motion $\{B_t\}$, in terms of which the information process $\{\xi_t\}$ is defined, is not adapted 
to the market filtration; and thus we cannot quite say that the asset price is ``driven'' by 
$\{B_t\}$ in the usual sense. 

To overcome this problem we make use of an idea from filtering theory---the idea of a so-called 
innovations process. In particular, we define a process $\{W_t\}$ by 
\begin{eqnarray}
W_t =  \xi_t -  \int^{t}_0 \mathbb{E}[X| \,{\mathcal F}_s]\rd s. 
\label{innovations}
\end{eqnarray}
One can show, for example, by use of the L\'evy criterion (see Appendix~\ref{app:2})
that $\{W_t\}$ is an $\{{\mathcal F}_t\}$-Brownian motion. Next we define a 
process $\{\lambda_t\}$ by setting
\begin{eqnarray}
\lambda_t = \mathbb{E}[X| \,{\mathcal F}_t].
\label{conditional exp of X}
\end{eqnarray}
By virtue of the relations (\ref{conditional expectation}), (\ref{innovations}), and 
(\ref{conditional exp of X}), the information process $\{\xi_t\}$ evidently satisfies 
a stochastic differential equation of the form 
\begin{eqnarray}
\rd \xi_t =  \lambda_t \rd t +  \rd W_t,
\label{SDE for xi}
\end{eqnarray}
where
\begin{eqnarray}
\lambda_t = 
\frac{\int^{\infty}_0 x \exp\left(x\xi_t-\tfrac{1}{2}x^2 t\right)p(\rd x)}
{\int^{\infty}_0 \exp\left(x\xi_t-\tfrac{1}{2}x^2 t\right)p(\rd x)},
\label{formula for lambda}
\end{eqnarray}
and it follows that the dynamical equation for the price can be put in the desired form 
\begin{eqnarray}
\rd S_t = (r  + \lambda_t\, \sigma ) S_t \,\rd t + \sigma S_t \, \rd W_t.
\label{SDE for price}
\end{eqnarray}

We note that the resulting ``effective'' market price of risk $\{\lambda_t\}$ is given by the 
conditional expectation (\ref{conditional exp of X}), and hence can be interpreted as the 
best estimate, given the 
information available, of the ``true'' value of the random variable $X$.  A little reflection 
shows that we can drop the adjective ``effective'' and simply assert that $\{\lambda_t\}$ is indeed 
the market price of risk (or market risk 
aversion level) in this model and that $X$ is the (unknown) \textit{actual} excess rate of 
return per unit of risk. Market participants acknowledge that only ``the gods'' know 
what the actual excess rate of return will turn out to be, or to have been, but that $\{\lambda_t\}$, 
which is 
knowable, represents the market consensus, the best estimate, the weighted opinion of 
market experts, the vote. \textit{The observable drift of an asset is thus determined not by the actual risk premium, but 
rather by the market best estimate for the risk premium}. In particular, given the price $S_t$ 
of the asset at time $t$, one deduces by use of (\ref{asset price xi}) and  (\ref{formula for lambda}) that the best estimate of the market price of risk is 
given by the following expression: 
\begin{eqnarray}
\lambda_t = 
\frac{\int^{\infty}_0 x (S_t/S_0)^{x/\sigma} \exp\left[-\tfrac{1}{2}x^2 t+(\frac{1}{2}\sigma - r/\sigma) xt 
\right]p(\rd x)}
{\int^{\infty}_0 (S_t/S_0)^{x/\sigma} \exp\left[-\tfrac{1}{2}x^2 t+(\frac{1}{2}\sigma - r/\sigma) xt \right]p(\rd x)}. 
\label{formula for lambda2}
\end{eqnarray}
This formula shows explicitly how the investor is able to update the \textit{a priori} 
estimate for the market price of risk given the current price level of the risky asset. 

It follows from (\ref{SDE for xi}) that the information 
process $\{\xi_t\}$ is a Brownian motion under the risk-neutral measure 
${\mathbb Q}$. Furthermore, if we make use of the market price of risk to effect a change of measure, a 
calculation shows that (a) 
the random variables $\xi_t$ and $X$ are independent under ${\mathbb Q}$, and (b) the 
probability law for $X$ under ${\mathbb Q}$ is given by $p(\rd x)$; that is to say, it is the same 
as it is under the physical measure ${\mathbb P}$. Therefore, an ``observer'' in the 
risk-neutral frame of reference $({\mathit\Omega}, {\mathcal F}, {\mathbb Q})$ detects the 
``message'' 
$\{\xi_t\}$, or equivalently the price $\{S_t\}$, but finds that it contains no information about 
the level of risk aversion---this is the sense in which the level of risk aversion cannot be 
inferred \textit{a priori} from derivative prices in the context of Brownian-motion based models. If stronger modelling assumptions are made about the structure of the pricing kernel in a Brownian model, then in 
some contexts it is possible to infer information about the risk aversion level from derivative 
prices (Andruszkiewicz \& Brody 2011, Ross 2011, Brody, Hughston \& Mackie 2012, 
Carr \& Yu 2012). The approach that we are taking is, perhaps, more practically oriented,  inasmuch as
an explicit estimation formula for the risk premium, such as that given by (\ref{formula for lambda2}), can 
be obtained in a direct and transparent manner without any reference to the risk-neutral measure. 

The $\{{\mathcal F}_t\}$-dynamics of the pricing kernel can be pursued similarly. In fact, 
it is more convenient first to work out the dynamics of the natural numeraire. Starting with 
equation (\ref {natural numeraire}), by a direct  application of Ito calculus we obtain
\begin{eqnarray}
\rd n_t = r  n_t \, \rd t +  \int^{\infty}_0 x 
\exp\left(rt + x\xi_t-\tfrac{1}{2}x^2 t\right)p(\rd x)\, \rd \xi_t .
\end{eqnarray}
It follows then by use of (\ref{SDE for xi}) and  (\ref{formula for lambda}) that
\begin{eqnarray}
\rd n_t = (r  + \lambda^{\,2}_t  ) \, n_t \,\rd t + \lambda_t  n_t \, \rd W_t, 
\end{eqnarray}
and therefore that
\begin{eqnarray}
\rd \pi_t = -r  \, \pi_t \,\rd t - \lambda_t  \pi_t \, \rd W_t .
\label{pricing kernel SDE}
\end{eqnarray}
Hence we are led to the conclusion that the volatility of the pricing kernel is (minus) the 
market price of risk, as it of course should be in the  $\{{\mathcal F}_t\}$-dynamics of the 
pricing kernel, and we can write
\begin{eqnarray}
\pi_t =  \exp \left ( -rt - \int^{t}_0  \lambda_s \, \rd W_s -  \int^{t}_0  \lambda_s^{\,2} \, \rd s \right ).
\end{eqnarray}
Such an expression for the pricing kernel or state price density is often used as the starting point of various investigations in the theory of finance---but  note that we have not assumed that $\{\pi_t\}$ takes this form, we have deduced it. 

As a model constructed in the spirit of classical mathematical finance, without 
direct mention of the risk aversion variable $X$, one thus has the following. We begin with 
a probability space $({\mathit\Omega},{\mathcal F},{\mathbb P})$, on which a standard 
Brownian motion $\{W_t\}$ is defined, and we let $\{{\mathcal F}_t\}$ be the associated 
filtration. The model inputs include the initial price $S_0$, the interest rate $r$, the volatility 
$\sigma$, and a measure $p(\rd x)$ on ${\mathds R}^+$.  With this data at hand, one 
defines a smooth function of two variables 
$\lambda : {\mathds R} \times {\mathds R}^+ \rightarrow {\mathds R}^+$  given by
$(\xi, t)  \rightarrow \lambda(\xi, t)$, where
\begin{eqnarray}
\lambda(\xi, t) = 
\frac{\int^{\infty}_0 x \exp\left(x\xi-\tfrac{1}{2}x^2 t\right)p(\rd x)}
{\int^{\infty}_0 \exp\left(x\xi-\tfrac{1}{2}x^2 t\right) p(\rd x)}.
\label{lambda function}
\end{eqnarray}
One can check that for fixed $t$ the function $\lambda(\xi, t)$ is increasing in the variable $\xi$. 
In particular, a calculation shows that $\lambda'(\xi, t)>0$, where the dash denotes differentiation 
with respect to  $\xi$. The process $\{\xi_t\}$ is then defined as the solution to the stochastic 
differential equation
\begin{eqnarray}
\rd \xi_t =  \lambda(\xi_t, t) \rd t +  \rd W_t,
\label{SDE2 for xi}
\end{eqnarray}
with the initial condition $\xi_0 = 0$. Having obtained $\{\xi_t\}$, one defines the process 
$\{\lambda_t\}$ by setting $\lambda_t = \lambda(\xi_t, t)$. The SDE 
for the asset 
price is taken to be (\ref{SDE for price}), with the initial condition $S_0$, and the 
SDE for the pricing kernel is taken to be given by (\ref{pricing kernel SDE}), with 
the initial condition unity. That gives a complete characterisation of the dynamics of the asset 
price and the pricing kernel. 

Having constructed the model in the filtration $\{{\mathcal F}_t\}$ without reference to the random variable $X$, one might ask whether it 
is possible in some sense to reconstruct $X$. It turns out that one can. In fact, we can \textit{derive} expressions for the two ``hidden'' objects $X$ and 
$\{B_t\}$ appearing in (\ref{information process}) from the ingredients arising in the $\{{\mathcal F}_t\}$ version of the modelling framework. Specifically, let $\lambda_t=\lambda(\xi_t,t)$, and let $\{\xi_t\}$ satisfy the stochastic differential equation  (\ref{SDE2 for xi}), with $\xi_0 = 0$. We can then show (i) that the 
random variables defined by 
\begin{eqnarray}
X=\lim_{T\to\infty}  T^{-1} \xi_T \qquad {\rm and} \qquad B_t=\xi_t- X t 
\label{random variables}
\end{eqnarray}
are independent for all $t$, (ii) that the distribution of $X$ is given by
$p(\rd x)$, and (iii) that 
the process $\{B_t\}$ thus arising is a standard ${\mathbb P}$-Brownian motion. The details of the arguments involved in establishing these facts are summarised in Appendix~\ref{app:3}.

%%%%%%%%%%%%%%%%%
%Section 5. Geometric L\'evy models
%%%%%%%%%%%%%%%%%
\section{Geometric L\'evy models} 
\label{sec:5}

\noindent Remarkably, the considerations that we have presented in connection with GBM models  generalise very naturally to the context of geometric L\'evy models (GLMs). As a consequence, we are able to construct a large and rich family of financial models for asset prices with jump risk in situations where the market exhibits variation among its participants in the excess rate of return required (above the interest rate) as compensation for the assumption of such risk.  We assume
familiarity with basics of the theory and application of L\'evy processes in what follows, as discussed 
for example in Cont \& Tankov (2004), 
Kyprianou (2006), Protter (1990), or Schoutens (2004). Numerous investigations have been pursued concerning the development of L\'evy-based  models in finance, and as a consequence the literature is very extensive. We mention for example the work of
Madan \& Seneta (1990),  
Madan \& Milne (1991), 
Heston (1993),
Gerber \& Shiu (1994), 
Eberlein \& Keller (1995), 
Eberlein \& Jacod (1997), 
Madan \textit {et al.}~(1998),
Chan (1999), 
 Carr \textit {et al.}~(2002), 
 Kallsen \& Shiryaev (2002), 
Hubalek \& Sgarra  (2006),
 Baxter (2007), and Yor (2007). To set the notation we begin with a few definitions.  A L\'evy process on a probability space 
$({\mathit \Omega},{\mathcal F},{\mathbb P})$ is a process $\{X_t\}$ such that $X_0=0$, 
$X_t-X_s$ is independent of ${\mathcal F}_s$ for $t\geq s$ (independent 
increments), and 
\begin{eqnarray}
{\mathbb P}(X_t-X_s\leq y) = {\mathbb P}(X_{t+h}-X_{s+h}\leq y) 
\end{eqnarray}
(stationary increments), 
where $\{{\mathcal F}_t\}$ denotes the augmented filtration generated by $\{X_t\}$. 
In order for $\{X_t\}$ to give rise to a geometric L\'evy model, we require that it should satisfy
\begin{eqnarray}
{\mathbb E}[\re^{\alpha X_t}] < \infty  
\end{eqnarray}
for all $t\geq 0$, for some connected real interval $\alpha\in A$ 
containing the origin.
We consider L\'evy processes satisfying such a condition. 
It follows then by the stationary and independent increments property that there exists 
a function $\psi(\alpha)$, the L\'evy exponent, such that 
\begin{eqnarray}
{\mathbb E}[\re^{\alpha X_t}] = \re^{t\psi(\alpha)}  
\label{cumulant}
\end{eqnarray}
for $\alpha \in \{w \in {\mathds C} : {\rm Re} (w) \in A \}$,
and one can check that the process
defined by
\begin{eqnarray}
M_t = \re^{\alpha X_t-t\psi(\alpha)}   
\label {martingale}
\end{eqnarray}
is an $\{{\mathcal F}_t\}$-martingale. We call $\{M_t\}$ the 
geometric L\'evy martingale (or Esscher martingale) associated with $\{X_t\}$, with 
parameter $\alpha$. For example, in the case of a standard Brownian motion the L\'evy exponent is given by
\begin{eqnarray}
\psi(\alpha)= \half \alpha^2,
\end{eqnarray}
which is defined for all real $\alpha$, and the associated geometric L\'evy martingale is a 
compensated geometric Brownian motion with volatility $\alpha$.  A L\'evy process is fully 
characterised by its exponent. As a consequence, it is useful to define and classify such 
processes by presentation of their L\'evy exponents. For example, the Poisson process 
with rate $m$ is defined by 
\begin{eqnarray}
\psi(\alpha) = m(\re^\alpha -1).
\end{eqnarray}
The gamma process with rate $m$ and scale unity is given by
\begin{eqnarray}
\psi(\alpha) = -m \ln(1-\alpha),
\label{psi gamma}
\end{eqnarray}
for $\alpha < 1$. The variance-gamma (VG) process is given for $\alpha^2 < 2m$ by
\begin{eqnarray}
\psi(\alpha) = -m \ln \left(1-\frac {\alpha^2}{2m}\right).
\end{eqnarray}

The theory of Esscher transformations plays an important role in the analysis of the relation between risk and return in L\'evy models. We say that two L\'evy exponents $\psi(\alpha)$ and $\tilde \psi(\alpha)$ are related by an Esscher transformation with parameter $\lambda$ if
\begin{eqnarray}
\tilde \psi(\alpha) = \psi(\alpha + \lambda) - \psi(\lambda).
\end{eqnarray}
The parameter $\lambda$ must lie in the domain $A$ of $\psi(\alpha)$, and the domain of $\tilde \psi(\alpha)$ consists of those values of $\alpha$ such that $\alpha + \lambda \in A$. The operation is reversible in the sense that if $\tilde \psi(\alpha)$ is an Esscher transformation of $\psi(\alpha)$ with parameter $\lambda$, then $\psi(\alpha)$ is an Esscher transformation of $\tilde \psi(\alpha)$ with parameter $-\lambda$. Thus if $\psi(\alpha) = \half \alpha^2$ represents a Brownian motion, then $\tilde \psi(\alpha) = \half \alpha^2 + \alpha \lambda$ represents a Brownian motion with drift $\lambda$. If $\psi(\alpha) = m(\re^\alpha -1)$ represents a Poisson process with rate $m$, then
$\tilde \psi(\alpha) = m\re^\lambda(\re^\alpha -1)$ represents a Poisson process with rate $m \re^\lambda$, and so on. Each example is rather different in character. 

With these definitions at hand, we can present a development of the theory of geometric L\'evy models for asset pricing (Brody, Hughston \& Mackie 2012) that is particularly well adapted to the analysis of jump risk aversion. 
The straightforward approach to geometric L\'evy models is as follows. First we 
construct the pricing kernel $\{\pi_t\}_{t\geq0}$. Let $\{X_t\}$ be a L\'evy process with exponent $\psi(\alpha)$ where $\alpha \in A$. Let $\lambda>0$ and assume that 
$-\lambda\in A$, and set 
\begin{eqnarray}
\pi_t = \re^{-rt} \re^{-\lambda X_t - t\psi(-\lambda)}.
\label{levy pricing kernel}
\end{eqnarray}
We require that the product of the pricing kernel and 
the asset price should be a martingale, which we assume is of the geometric L\'evy form
\begin{eqnarray}
\pi_tS_t = S_0\re^{\beta X_t - t\psi(\beta)}
\end{eqnarray}
for some $\beta\in A$, and we deduce that 
\begin{eqnarray}
S_t = S_0 \, \re^{rt}\, \re^{\sigma X_t+t\psi(-\lambda)-t\psi(\sigma-\lambda)},
\label{pricex}
\end{eqnarray}
where $\sigma = \beta + \lambda$. We shall assume that  $\sigma > 0$ and that 
$\sigma\in A$. It follows that the price can be expressed 
by the formula
\begin{eqnarray}
S_t = S_0 \, \re^{rt}\, \re^{R(\lambda,\sigma)t}\, \re^{\sigma X_t-t\psi(\sigma)},
\end{eqnarray}
where
\begin{eqnarray}
R(\lambda,\sigma) = \psi(\sigma) + \psi(-\lambda) - \psi(\sigma-\lambda). 
\label{R}
\end{eqnarray}
It is a remarkable fact that the excess rate of return function $R(\lambda,\sigma)$ thus arising is positive and is increasing with respect to both of its arguments.

%%%%%%%%%%%%%%%%%
%Section 6. On the aggregation of jump-risk aversion
%%%%%%%%%%%%%%%%%
\section{On the aggregation of jump-risk aversion} 
\label{sec:6}

\noindent In our analysis of jump-risk aversion, it will be useful to cast the foregoing formulation of geometric L\'evy models 
into a slightly different form which turns out to be well suited for our purpose. Let $\{X_t\}$ and 
$\psi(\alpha)$ be defined as above, and set 
\begin{eqnarray}
\phi(\alpha) = \psi(\alpha - \lambda) - \psi(-\lambda). 
\end{eqnarray}
Clearly we have
\begin{eqnarray}
\psi(\alpha) = \phi(\alpha + \lambda) - \phi(\lambda), 
\end{eqnarray}
and we observe that $\psi(\alpha)$ is given by an Esscher transform of $\phi(\alpha)$, with 
parameter $\lambda$. An exercise then shows that the asset price (\ref{pricex}) can be 
expressed in the form 
\begin{eqnarray}
S_t = S_0 \, \re^{rt}\, \re^{\sigma X_t-t\phi(\sigma)}.
\label{submartingale}
\end{eqnarray}
Thus we see that given a ``fiducial'' L\'evy exponent $\phi(\alpha)$, if we let the exponent of the 
L\'evy process $\{X_t\}$ be an Esscher transform of $\phi(\alpha)$, with parameter $\lambda$, 
then the process $S_t$ defined as above by (\ref{submartingale}) will be a submartingale, which we 
can take to be the asset price process. The associated pricing kernel is then a supermartingale that takes the form
\begin{eqnarray}
\pi_t = \re^{-rt} \re^{-\lambda X_t + t\phi(\lambda)}.
\label{levy pricing kernel2}
\end{eqnarray}
The advantage of this representation of the geometric L\'evy model (which is entirely equivalent to that of the previous section) is that it readily 
generalises to the situation where the risk aversion parameter is uncertain, thus allowing one to generalise the scheme set out in Sections I - IV to models with jumps.

To proceed further it will be expedient to make use of the general filtering theory associated with L\'evy 
noise developed in Brody, Hughston \& Yang (2013). We recall that, when phrased in the language of signal processing, 
what amounts to the ``signal'' in the present investigation is the unknown level of risk aversion. In 
the Brownian context, it is natural for the signal to be obscured by an additive noise. However, in 
the case of general L\'evy noise with jumps, the signal is no longer obscured by noise in an additive 
manner. In fact, each different type of L\'evy process, when viewed as a model for noise, ``carries'' the signal in its 
own distinctive manner. 

Rather than developing the theory case by case, here we propose to present the theory in such a way that is 
applicable to the whole category of L\'evy models. For this purpose we introduce the important notion of  
\textit {L\'evy information}. 
By a L\'evy information process $\{\xi_t\}$ with signal $X$ on a probability space 
$({\mathit \Omega},{\mathcal F},{\mathbb P})$, we mean a process 
such that conditional on the sigma field ${\mathcal F}^X$ generated by $X$, $\{\xi_t\}$ is a 
L\'evy process with exponent 
\begin{eqnarray}
\psi_X(\alpha) = \phi(\alpha + X) - \phi(X)  
\end{eqnarray}
for $\alpha \in \{w \in {\mathds C} : {\rm Re} (w) = 0\}$. That is to say, we have the relation 
\begin{eqnarray}
{\mathbb E}\left[ \re^{\alpha \xi_t}| {\mathcal F}^X \right] = \re^{t\psi_X(\alpha)}  
\end{eqnarray}
for imaginary values of $\alpha$. In terms of the L\'evy information process $\{\xi_t\}$, the asset price can be expressed in the form 
\begin{eqnarray}
S_t = S_0 \re^{rt} \re^{\sigma \xi_t - t \phi(\sigma)},
\label{eq:51}
\end{eqnarray}
which can be thought of as the general L\'evy-based analogue of the price process (\ref{asset price xi}). 
One way of looking at the formula above is to view $\{\xi_t\}$ as the driving L\'evy process in the risk-neutral measure 
${\mathbb Q}$, with respect to which the associated L\'evy exponent is given by $\phi(\alpha)$. Under ${\mathbb Q}$, however, $\{\xi_t\}$ encodes no information 
about the level of risk aversion.  For this we need to work with the physical measure ${\mathbb P}$ and identify the pricing 
kernel. A naive candidate for $\{\pi_t\}$, analogous to 
(\ref{eq:14}), is given by the formula
\begin{eqnarray}
\pi_t \stackrel{?}{=}  \re^{-rt} \re^{-X \xi_t + t\phi(X)}. 
\end{eqnarray}
This expression suffers 
from the fact that it is not measurable with respect to the sigma field ${\mathcal F}_t$ generated 
by the trajectory of the information process up to time $t$. It turns out, fortunately, and perhaps surprisingly, that the approach taken in the case of the 
Brownian example carries through to the general L\'evy context, and for the pricing kernel we 
have: 
\begin{eqnarray}
\pi_t = \mathbb{E}_t \big[\exp\left( -rt  -X \xi_t + \phi(X)t \right) \big] =  
\int^{\infty}_0  \exp\left( -rt  -x \xi_t + \phi(x) t \right) \, p_t(\rd x),
\label{eq:52}
\end{eqnarray}
where 
\begin{eqnarray}
p_t(\rd x) = 
\frac{\exp\left(x\xi_t-\phi(x)t\right)p(\rd x)}
{\int^{\infty}_0 \exp\left(z\xi_t-\phi(z)t\right) p(\rd z)}
\label{eq:53}
\end{eqnarray}
is the $\{{\mathcal F}_t\}$-conditional measure for the distribution of $X$, and $p(\rd x)$ is the unconditional measure. Substituting (\ref{eq:53}) in (\ref{eq:52}) we 
thus deduce that the pricing kernel in a geometric L\'evy model with random risk aversion is given 
by the following formula:
\begin{eqnarray}
\pi_t = \frac{1}{\int^{\infty}_0 \exp\left( rt + x\xi_t-\phi(x)t\right)p(\rd x)}.
\label{eq:54}
\end{eqnarray}
Similarly, for the natural numeraire in the case of a L\'evy model with random risk aversion we can write  
\begin{eqnarray}
n_t = \int^{\infty}_0 \exp\left( rt + x\xi_t-\phi(x)t\right)p(\rd x).
\label{eq:55}
\end{eqnarray}
Again, as in the Brownian situation, one observes in the L\'evy case the key point that the market numeraire asset can be viewed as
a portfolio, the elements of which correspond, with appropriate weights, to the numeraire assets of the various investors, each with a volatility given by the risk aversion factor associated to the particular investor.

With the conditional density (\ref{eq:53}) at hand we are able to determine the optimal estimate (in the sense 
of least quadratic error) for the level of jump-risk aversion. This is given by  
\begin{eqnarray}
\lambda_t = 
\frac{\int^{\infty}_0 x \exp\left(x\xi_t-\phi(x) t\right)p(\rd x)}
{\int^{\infty}_0 \exp\left(x\xi_t-\phi(x) t\right) p(\rd x)}.
\end{eqnarray}
Since the asset price (\ref{eq:51}) is a simple invertible function of $\xi_t$ we are thus in a position to obtain 
an explicit formula for the jump-risk aversion factor $\lambda_t$ in terms of the price level $S_t$ in the general setting of a
geometric L\'evy model. Once $\lambda_t$ has been determined, the excess rate of return associated with 
jump risk is given by $R(\lambda_t,\sigma)$. 
It is perhaps remarkable that the analysis presented in the case of the geometric Brownian motion model extends 
so straightforwardly to the case of the general geometric L\'evy model, even though the powerful tools of the 
traditional Ito calculus are not directly applicable in the general L\'evy context. This can be viewed 
as a vindication of the usefulness of  pricing kernel methods. Indeed, to grasp the relation 
between risk, risk aversion, and return, the pricing kernel is an indispensable tool, and this is especially clear when 
prices can jump, as is in any event typically the case in real financial markets. In particular, since optimal investment strategies depend solely for their specification on the pricing kernel and the risk profile of the investor (as given, for example, by an appropriate utility function), we are led by this reasoning to be able to present a clear account of such strategies in the case of markets with price jumps.

%%%%%%%%%%
%Acknowledgements
%%%%%%%%%%
\begin{acknowledgments}
\noindent The authors are grateful to I. R. C. Buckley, M. R. Grasselli, E. Mackie, X. Yang, J. P. Zubelli, and participants in the 2012 Research in Options  conference in B\'uzios, Rio de Janeiro, where a preliminary version of this work was presented, for helpful comments.
\end{acknowledgments}

%%%%%%%%%%
%REFERENCES
%%%%%%%%%%

\vskip 15pt \noindent {\bf References}.
%+Bibliography

%%%%%%%
%Appendix A
%%%%%%%
 \appendix%{} 

\section{Conditional Distribution of the Risk Aversion Factor} 
\label{app:1} 

\noindent 
In this appendix we present the details of a calculation leading to the general expression given by equation (\ref{conditional distribution}) for the conditional distribution of risk aversion factor in the case of a filtration generated by a Brownian information process.

We consider a probability space $({\mathit\Omega},{\mathcal F},{\mathbb P})$  on which a Brownian motion $\{B_t\}_{t\geq 0}$ is defined along with an independent square-integrable random variable $X$, and we assume that $X>0$ almost surely. The filtration $\{{\mathcal F}_t\}$ is taken to be generated by the information process $\{\xi_t\}_{t\geq 0}$ defined by $\xi_t=B_t+Xt$.
The conditional distribution of $X$ can be worked out as follows. 

First, we note (i) that $\{\xi_t\}$ is a Markov process
and (ii) that $X$ is ${\mathcal F}_\infty$-measurable. To establish the Markov property we note the fact that in the case of Brownian motion the random variables $B_t$ and $B_s/s-B_{s_1}/s_1$ are independent for $t>s>s_1>0$ by virtue of the theory of the Brownian bridge.  
More generally, if $s>s_1>s_2>s_3>0$, we find that
$B_s/s-B_{s_1}/s_1$ and $B_{s_2}/s_2-B_{s_3}/s_3$ are independent.  
We observe that for any $k \geq 1$ we have
\begin{eqnarray}
{\mathbb P}\left( \xi_t\leq x| \xi_s,\xi_{s_1},
\ldots,\xi_{s_k}\right) &=& {\mathbb P}\left(\xi_t\leq x\Big| \xi_s,
\frac{\xi_s}{s}-\frac{\xi_{s_1}}{s_1}, \ldots, \frac{\xi_{s_{k-1}}}{s_{k-1}} -
\frac{\xi_{s_k}}{s_k}\right) \nonumber \\ &=& 
{\mathbb P}\left( \xi_t\leq x\Big| \xi_s, \frac{B_s}{s}-
\frac{B_{s_1}}{s_1}, 
\ldots, \frac{B_{s_{k-1} }}{s_{k-1}}-\frac{B_{s_k}}{s_k}\right).
\end{eqnarray}
Since $\xi_t$ and $\xi_s$ are independent of $B_s/s-B_{s_1}/s_1$, 
$\ldots$, $B_{s_{k-1}}/s_{k-1}-B_{s_k}/s_k$, it follows that 
\begin{eqnarray}
{\mathbb P}\left( \xi_t\leq x| \xi_s,\xi_{s_1},\ldots,
\xi_{s_k}\right) = {\mathbb P}\left( \xi_t\leq x|\xi_s\right),
\end{eqnarray}
and that gives us the Markov property. As regards the ${\mathcal F}_\infty$-measurability, 
this follows from the fact that $\lim_{t\to\infty}t^{-1}\xi_t=X$. In the calculation of the conditional distribution of $X$ given  ${\mathcal F}_t$ it thus suffices to determine 
the conditional distribution of $X$ given $\xi_t$. By virtue of the relevant version of the Bayes formula 
we have 
\begin{eqnarray}
p_t(\rd x) = \frac{\rho(\xi_t|X=x) p(\rd x) }{\int_0^\infty 
\rho(\xi_t|X=x)p(\rd x)}, \label{eq:08}
\end{eqnarray}
where $p(\rd x) = {\mathbb P}(X \in \rd x)$ is the {\it a priori} distribution of $X$, assumed known, and where for each $x$ the function $\rho(\xi|X=x)$, $\xi \in \mathbb R$,  is the conditional density for the random 
variable $\xi_t$ given that $X=x$, which in (\ref{eq:08}) is then valued at $\xi = \xi_t({\mathit  \omega})$ for each outcome of chance ${\mathit \omega} \in {\mathit \Omega}$. Since $B_{t}$ is a Gaussian random variable with 
mean $0$ and variance $t$, we deduce that
\begin{eqnarray}
\rho(\xi|X=x) = \frac{1}{\sqrt{2\pi t}} \, 
\exp\left( -\frac{(\xi- t x)^2}{2t}\right) .
\label{eq:4.13}
\end{eqnarray}
Inserting this expression into the Bayes formula (\ref{eq:08}), one is then immediately led to (\ref{conditional distribution}). 

%Appendix 2 Emergence of the Brownian Driver%
\section{Emergence of the Brownian Driver} 
\label{app:2} 

\noindent 
In the conventional modeling framework (and in the absence of jumps apart from those associated directly with dividend payments) it is usually assumed that the market filtration is generated by a Brownian motion of one or more dimensions, and that the associated asset prices are adapted to this filtration. Although well-established and mathematically sound, from a financial perspective this view of the market is unsatisfactory in various respects. One gets a hint at the nature of the problem when on the one hand (a) the Brownian motion $\{W_t\}$ driving the asset is referred to as  ``noise'', and on the other hand (b) the sigma field
${\mathcal F}_t = \sigma[\{W_s\}_{0\leq s \leq t}]$ is referred to as ``information''. If one presses a finance theorist on this point, the reply will be a shrug of the shoulders and piece of sophistry of the form, ``Well, it is true that  $\{W_t\}$ is noise, but  ${\mathcal F}_t$ represents the knowledge of the history of the trajectory that noise, and therefore carries valuable information.'' This point of view, ridiculous as it may seem, permeates the whole subject, and is consequently a source of confusion. 

Fortunately, there is a resolution of this seemingly paradoxical issue. Prices are driven by the flow of information, and information is usually communicated along noisy channels. The market accepts this as the normal state of affairs, and prices are based on the best estimates of the relevant factors, given the information available, imperfect as it may be. This is perhaps what Norbert Wiener was getting at when he said that ``Economics is a science of communication'' (Jerison \& Stroock 1997). 

In this appendix we present details of the calculations in Section IV leading to the emergence of the Brownian driver. Starting with the relation $\xi_t=B_t+Xt$, we define the process $\{W_t\}$ as in (\ref{innovations}).
To prove that $\{W_t\}$ is an $\{{\mathcal F}_t\}$-Brownian motion it suffices by use of the so-called L\'evy criterion to show 
that $\{W_t\}$ is an $\{{\mathcal F}_t\}$-martingale and that $(\rd W_t)^2=\rd t$. First, 
we shall demonstrate that $\{W_t\}$ is an $\{{\mathcal F}_t\}$-martingale. Letting 
$t\leq T$ we deduce that
\begin{eqnarray}
{\mathbb E}\left[W_T|{\mathcal F}_t\right]&=& 
%{\mathbb E}
%\left[\xi_T| {\mathcal F}_t\right]- {\mathbb E}
%\left[\left. \int_0^T \lambda_s {\rm d}s\right| {\mathcal F}_t\right]
%\nonumber \\ 
%&=& 
{\mathbb E}\left[ B_T| {\mathcal F}_t\right] + T {\mathbb E}\left[ X| {\mathcal F}_t\right] 
- {\mathbb E} \left[\left. \int_0^T 
\lambda_s {\rd}s\right| {\mathcal F}_t\right] \nonumber \\ &=& 
{\mathbb E}\left[ B_T| {\mathcal F}_t\right] + T {\mathbb E}\left[ X| {\mathcal F}_t\right]  - \int_0^T
{\mathbb E} \left[ \lambda_s | {\mathcal F}_t\right]{\rd}s,
\label{eq:96}
\end{eqnarray}
by use of Fubini's theorem. Next, we note that
\begin{eqnarray}
\int_0^T {\mathbb E} \left[ \lambda_s | {\mathcal F}_t\right]
{\rm d}s =
\int_0^t {\mathbb E} \left[ \lambda_s | {\mathcal F}_t 
\right]{\rd}s + \int_t^T {\mathbb E} \left[ \lambda_s |
{\mathcal F}_t\right]{\rd}s 
%\\ &=& \int_0^t \lambda_s
%{\rd}s + \int_t^T \lambda_t {\rm d}s \nonumber 
= \int_0^t \lambda_s
{\rd}s + (T-t) \lambda_t . \label{eq:97}
\end{eqnarray}
Here we have used the fact that the process $\{\lambda_t\}$ is by construction an $\{{\mathcal F}_t\}$-martingale. 
Substituting (\ref{eq:97}) in (\ref{eq:96}) we obtain
\begin{eqnarray}
{\mathbb E}\left[W_T|{\mathcal F}_t\right] =   {\mathbb E}
\left[ B_T| {\mathcal F}_t\right] + t
{\mathbb E}\left[ X| {\mathcal F}_t\right] - \int_0^t \lambda_s
{\rd}s . \label{eq:98}
\end{eqnarray}
Finally, we observe that by the tower property of conditional
expectation we have
\begin{eqnarray}
{\mathbb E}\left[B_T|{\mathcal F}_t\right] = {\mathbb E}
\left[ {\mathbb E}\left[B_T| {\mathcal F}_t^B,X\right]
\Big|{\mathcal F}_t \right] = {\mathbb E}\left[B_t|
{\mathcal F}_t \right] , \label{eq:99}
\end{eqnarray}
where $\{{\mathcal F}_t^B\}$ denotes the filtration generated by $\{B_t\}$. 
Inserting this in (\ref{eq:98}) we obtain
\begin{eqnarray}
{\mathbb E}\left[W_T|{\mathcal F}_t\right] &=&  {\mathbb E}\left[ 
B_t| {\mathcal F}_t\right] + t \lambda_t  -\int_0^t \lambda_s {\rd}s \nonumber \\ &=& {\mathbb E} 
\left[(B_t+ t X )| {\mathcal F}_t\right] - \int_0^t \lambda_s {\rd}s \nonumber \\ &=& 
{\mathbb E}\left[ \xi_t| {\mathcal F}_t\right] -\int_0^t \lambda_s \rd s = W_t \ , 
\label{eq:100}
\end{eqnarray}
and this establishes that $\{W_t\}$ is an $\{{\mathcal F}_t\}$-martingale. Next, we 
observe that since
\begin{eqnarray}
{\rd}W_t=(X-\lambda_t){\rd}t+{\rd} B_t,
\end{eqnarray}
it follows at once that $({\rd}W_t)^2={\rm d}t$. Taking this result together with the fact that $\{W_t\}$ 
is an $\{{\mathcal F}_t\}$-martingale, we conclude that $\{W_t\}$ is an 
$\{{\mathcal F}_t\}$-Brownian motion.

%%%%%%%%%%%
%% Appendix C %%
%%%%%%%%%%%
\section{Existence and Construction of the Hidden Variables $X$ and $B_t$} 
\label{app:3} 

\noindent
In this appendix we present details of the arguments allowing one to establish properties (i), (ii), and (iii) of  the constructed versions of the random risk 
aversion variable $X$ and the associated ``pure noise'' process $\{B_t\}$ stated at the end of Section IV, starting from the formulation of the theory 
in which all quantities under consideration at the outset are ``financial observables'', that is to say, suitably adapted to the market filtration. Let us begin by establishing property (i), the independence of the random variables $B_t$ and $X$. 
To this end  it suffices to check that the relation 
\begin{eqnarray}
{\mathbb E}\left[\re^{a B_t+b X}\right]={\mathbb E}\left[\re^{a B_t}\right] 
{\mathbb E}\left[\re^{b X}\right]
\label{eq:11.1}
\end{eqnarray}
holds for all $a,b \in \mathds C^{\rm I} := \{w \in \mathds C : {\rm Re} \, w = 0\} $. Verification that the joint characteristic function factorizes  proceeds as follows. By the definitions of $B_t$ and $X$ given at (\ref{random variables}) we have  
\begin{eqnarray}
{\mathbb E}[\re^{aB_t+bX}] 
= \lim_{T\to\infty} {\mathbb E}[\re^{a (\xi_t - tT^{-1} \xi_T)+ b  T^{-1} \xi_T }].
 \label{limit of expectation}
\end{eqnarray}
We shall calculate the expectation in (\ref{limit of expectation}) and show that it factorizes for all $T$. In this connection it will be useful to construct a solution to the stochastic differential equation 
({\ref{SDE2 for xi}). We begin with a probability space 
$({\mathit\Omega}, {\mathcal F},{\mathbb Q})$ on which we introduce a standard Brownian motion $\{\xi_t\}$. Given the measure $p(\rd x)$, which we assume to admit a second moment, one can check that the function $\Phi(\xi, t)$ defined for $\xi, t \geq 0$ by 
\begin{eqnarray}
\Phi(\xi, t) = \int_0^\infty \exp\left(x \xi - \half x^2 t \right) p(\rd x) 
\end{eqnarray}
is of class $\rm C^2$ in $\xi$ and class $\rm C^1$ in $t$, and (Yor 1992) has the space-time harmonic property, 
\begin{eqnarray}
\frac {\partial \Phi} {\partial t} = \half \frac {\partial^2 \Phi} {\partial \xi^2}.
\end{eqnarray}
As a consequence 
we are able to introduce a process $\{\Phi_t\}$ defined by 
\begin{eqnarray}
\Phi_t = \Phi(\xi_t, t) = \int_0^\infty \exp\left(x \xi_t  - \half x^2 t \right) p(\rd x), 
\end{eqnarray}
and it is straightforward to verify that $\{\Phi_t\}$ is a martingale under ${\mathbb Q}$ with respect to the filtration $\{{\mathcal F}_t\}$ generated by  $\{\xi_t\}$. Applying Ito's lemma, and defining the process 
$\{\lambda_t\}$ as before by $\lambda_t = \lambda(\xi_t, t)$, where the function $\lambda(\xi, t)$
is given by $(\ref{lambda function})$, one deduces that 
$\rd  \Phi_t = \lambda_t \Phi_t \rd \xi_t$,
and hence by integration we obtain 
\begin{eqnarray}
\int_0^\infty \exp\left(x \xi_t  - \half x^2 t \right) p(\rd x) = \exp\left( 
\int_0^t \lambda_s \rd \xi_s - \half \int_0^t \lambda_s^2\rd s\right), 
\label{eq:11.4}
\end{eqnarray}
which expresses $\{\Phi_t\}$ in the form of an exponential martingale. Since $\{\xi_t\}$ is a ${\mathbb Q}$-Brownian motion, one sees by use of Girsanov's theorem that the process
$\{W_t\}$ defined by
\begin{eqnarray}
W_t = \xi_t - \int_0^t \lambda_s \rd s 
\label{W}
\end{eqnarray}
is a Brownian motion under the measure ${\mathbb P}$ defined by
\begin{eqnarray}
 \left. \frac{\rd{\mathbb P}} {\rd{\mathbb Q}}\right|_{{\mathcal F}_t} =  \Phi_t, 
\end{eqnarray}
and one concludes from (\ref {W}) that $\{\xi_t\}$ satisfies the stochastic differential equation ({\ref{SDE2 for xi}). We see moreover that ${\mathbb Q}$ is the risk-neutral measure.
The conditional expectations in the probability measures ${\mathbb P}$ and ${\mathbb Q}$ are related for $0 \leq t \leq T$ by 
the scheme
\begin{eqnarray}
{\mathbb E}_t^{\mathbb P}[Y_T]=\frac{1}{\Phi_t}{\mathbb
E}_t^{\mathbb Q} [\Phi_T Y_T] \quad {\rm and}\quad {\mathbb
E}_t^{\mathbb Q}[Y_T] = \Phi_t{\mathbb E}_t^{\mathbb P}
\left[\frac{1}{\Phi_T} Y_T\right] \label{eq:11.5}
\end{eqnarray}
for any 
${\mathcal F}_T$-measurable random variable 
$Y_T$. 

Equipped with these results we proceed to work out the expectation (under ${\mathbb P}$) appearing in
 (\ref{limit of expectation}). In particular, we need the relation 
 \begin{eqnarray}
 {\mathbb E}^{\mathbb P}[Y_T] = {\mathbb
E}^{\mathbb Q} [\Phi_T Y_T].
\end{eqnarray}
We thus observe that
 \begin{eqnarray}
 {\mathbb E}[\re^{a (\xi_t - tT^{-1} \xi_T)+ b T^{-1} \xi_T }] 
 &=& 
 {\mathbb E}^{\mathbb Q}\left[  
\left ( \int_0^\infty \re^{ x \xi_T  -\frac{1}{2} x^2 T} p(\rd x)\right ) \, \re^{a (\xi_t - tT^{-1} \xi_T)+ b  T^{-1} \xi_T }   \right] \nonumber \\ 
 &=& 
 \int_0^\infty  {\mathbb E}^{\mathbb Q}\left[  
\re^{ x \xi_T  -\frac{1}{2} x^2 T} \re^{a (\xi_t - tT^{-1} \xi_T)+ b T^{-1} \xi_T }  \right]  p(\rd x).
\end{eqnarray}
But since $\{\xi_t\}$ is a Brownian motion under 
${\mathbb Q}$, the inner expectation can be worked out by use of 
standard techniques from the theory of Brownian motion. 
The result is:
\begin{eqnarray}
{\mathbb E}^{\mathbb Q}\left[  
\re^{ x \xi_T  -\frac{1}{2} x^2 T}  \re^{a (\xi_t - tT^{-1} \xi_T)+ b  T^{-1} \xi_T }  \right]  
= \re^{ \frac{1}{2} a^2 t(T-t)T^{-1}} \re^{bx + \frac{1}{2} b^2 T^{-1} } , 
\label{eq:11.6} 
\end{eqnarray}
from which it follows that
\begin{eqnarray}
 {\mathbb E}[\re^{a (\xi_t - tT^{-1} \xi_T)+ b  T^{-1} \xi_T }] 
= \re^{ \frac{1}{2} a^2 t(T-t)T^{-1}}  \left(\int_0^\infty \re^{bx + \frac{1}{2} b^2 T^{-1}}\, p(\rd x) 
\right) , 
\end{eqnarray}
which exhibits the claimed factorization of the characteristic function for all $T$. In particular, for large $T$ we obtain
\begin{eqnarray}
 {\mathbb E}[\re^{a B_t + b  X }] 
= \re^{ \frac{1}{2} a^2 t}  \left(\int_0^\infty \re^{bx }\, p(\rd x) 
\right) , 
\end{eqnarray}
which establishes  (\ref{eq:11.1}), showing that random variables $X$ and $B_t$
defined by (\ref{random variables}) are independent for all $t$, which is property (i).  It follows further that the 
distribution of $X$ is given by $p(\rd x)$, which is property (ii), and that $B_t$ is normally distributed with mean zero and variance $t$, which gives us part of property (iii).
To complete the proof of property (iii), that $\{B_t\}$ is a Brownian motion, we must verify that 
$\{B_t\}$ has independent increments.
It will suffice to demonstrate that 
\begin{eqnarray}
{\mathbb E}\left[\re^{aB_t+b(B_u-B_t)}\right] = 
{\mathbb E}\left[\re^{aB_t}\right]{\mathbb E}\left[\re^{b(B_u-B_t)}\right]
\label{eq:11.8}
\end{eqnarray}
holds for $0 \leq t \leq u$ and $a,b \in \mathds C^{\rm I}$. 
Using the definition of $\{B_t\}$ we can write
\begin{eqnarray}
{\mathbb E}\left[\re^{aB_t+b(B_u-B_t)}\right] 
=
%&=&
%{\mathbb E}
%\left[   \re^{a (\xi_t - tX)+ b ((\xi_u - uX) - (\xi_t - tX)) }  \right] 
%\nonumber \\ &=&
{\mathbb E}^{\mathbb Q}
\left[ \Phi_T \, \re^{a (\xi_t - tX)+ b ((\xi_u - uX) - (\xi_t - tX)) }  \right], 
\end{eqnarray}
where $\Phi_T = \Phi(\xi_T, T)$, and it follows from the definition of $X$ that 
\begin{eqnarray}
{\mathbb E}\left[\re^{aB_t+b(B_u-B_t)}\right] 
=
%&=&
% \lim_{T\to\infty} {\mathbb E}^{\mathbb Q}
%\left[ \Phi_T \,  \re^{a (\xi_t - tT^{-1} \xi_T)+ b ((\xi_u - uT^{-1} \xi_T) - (\xi_t - tT^{-1} \xi_TX)) }  \right]  
%\nonumber \\ &=&
 \lim_{T\to\infty} {\mathbb E}^{\mathbb Q}
\left[  \re^{a (\xi_t - tT^{-1} \xi_T)+ b ((\xi_u - uT^{-1} \xi_T) - (\xi_t - tT^{-1} \xi_TX)) }  \right]. 
\label{ind inc}
\end{eqnarray}
In obtaining (\ref {ind inc}) we have used the theory of the Brownian bridge to deduce
that $\xi_T$ (and hence $\Phi_T$) is independent of $\xi_t - tT^{-1} \xi_T$
and $\xi_u - uT^{-1} \xi_T$ under ${\mathbb Q}$, and we have used the fact that ${\mathbb E}^{\mathbb Q} \left[ \Phi_T\right] = 1$. One is then left with a calculation involving the expectation 
of an exponentiated sum of Gaussian random variables, which can be simplified by use of the theory of the Brownian bridge, and for large $T$ we obtain the desired result:
\begin{eqnarray}
{\mathbb E}\left[\re^{aB_t+b(B_u-B_t)}\right]  
= \re^{ \frac{1}{2} a^2 t  }  \re^{ \frac{1}{2} b^2 (u -t) }.
\end{eqnarray}
The same line of argument applies for any number of increments.  Thus, we conclude that $\{B_t\}$ is normally distributed with zero mean and variance $t$, and has independent increments. Therefore, $\{B_t\}$ is a standard Brownian motion under ${\mathbb P}$.

\end{document}